\documentclass[amsmath,amssymb,prx,superscriptaddress,reprint,showpacs,longbibliography]{revtex4-2}
\usepackage{graphicx}
\usepackage{dcolumn}
\usepackage[colorlinks=true,linkcolor=blue,citecolor=blue,urlcolor=blue]{hyperref}
\usepackage[usenames,dvipsnames]{xcolor}
\usepackage{soul}
\usepackage{braket}
\usepackage{bm}
\usepackage{upgreek}
\usepackage{mathtools}
\usepackage{bbold}

\renewcommand{\vec}[1]{{\bf #1}}

\begin{document}

\title{Quantum Hall Superconductivity from Moir{\'e} Landau Levels}
\author{Gaurav Chaudhary}
\email{gchaudhary@anl.gov}
\affiliation{Materials Science Division, Argonne National Laboratory, Lemont, IL 60439, USA}
\author{A. H. MacDonald}
\affiliation{Department of Physics, The University of Texas at Austin, Austin,  TX 78712, USA}
\author{M. R. Norman}
\affiliation{Materials Science Division, Argonne National Laboratory, Lemont, IL 60439, USA}


\date{\today}

\pacs{}
\keywords{}

\begin{abstract}
It has long been speculated that quasi-two-dimensional
superconductivity can reappear above its semiclassical upper critical field due to Landau quantization, 
yet this reentrant property has never been observed.
Here, we argue that twisted bilayer graphene at a magic angle (MATBG) 
is an ideal system in which to search for this phenomenon because 
its Landau levels are doubly degenerate, and its superconductivity 
appears already at carrier densities small enough to allow the quantum limit
to be reached at relatively modest magnetic fields. 
We study this problem theoretically by combining a simplified continuum model for the electronic structure of 
MATBG with a phenomenological attractive pairing interaction, and discuss obstacles 
to the observation of quantum Hall superconductivity presented by 
disorder, thermal fluctuations, and competing phases.
\end{abstract}
\maketitle

\section{Introduction\label{Sec:Intro}}
Magnetic fields suppress superconductivity owing to 
either Pauli or orbital pair breaking, or a combination of the two.
Under most circumstances, superconductivity is not 
possible above an upper critical field $H_{c2\perp}$ that is small enough to justify
a weak-field semiclassical approximation.
However, as first proposed over fifty years ago, mean-field theory predicts that 
under the favorable circumstances specified below, Landau level (LL) degeneracy can cause 
superconductivity to reemerge in quantizing perpendicular magnetic fields~\cite{Gruenberg1968,Tesanovic1989,Rajagopal1991,Rasolt1992,MacDonald1992,MacDonald1993}. 
The predicted effect becomes particularly dramatic in two-dimensional systems with resolved low index 
doubly degenerate LLs. Although the theory of superconductivity in quantizing magnetic fields has been developed in great detail~\cite{Tesanovic1991,Norman1992,Dukan1997,Maska2002,Scherpelz2013,Song2017}, 
reemergence has never been observed.  We refer to this proposed state of matter as a quantum Hall superconductor.

Quantum Hall superconductivity 
requires near degeneracy between LLs that are 
distinguished by an internal label.  Because of Zeeman coupling, degeneracy between LLs with 
opposite spins can occur only when the orbital LL splitting fortuitously matches the Zeeman splitting.
Graphene bilayers are attractive candidates for quantum Hall  superconductivity in the first place because 
their LLs are labelled not only by spin, but also by $\pm$ valley indices, and are nearly 
valley degenerate unless aligned with the encapsulating hexagonal boron nitride 
(hBN) layers.  Given the valley degree of freedom, it is possible to draw pairs from degenerate low index
LLs with $\omega_c \tau \gg 1$ where $\hbar \omega_c$ is the 
LL separation and $\hbar/\tau$ is the LL width.
Because these prerequisites for quantum Hall superconductivity are rarely satisfied, twisted bilayer 
graphene provides a rare opportunity to 
pursue exotic quantum Hall pair states.
A number of proposals to engineer topological superconducting states rely on pairing of Landau quantized electrons~\cite{Zocher2016,Jeon2019,Chaudhary2020}.  The widespread interest in topological superconductivity and Majorana modes~\cite{Read2000,Ivanov2001,Alicea2012,Kitaev2003,Nayak2008} therefore adds 
motivation to quantum Hall superconductivity searches, beyond intrinsic interest in their novelty and 
their exotic vortex lattices~\cite{Akera1991}.

The theory that predicts reentrant quantum Hall 
superconductivity has not been fully tested 
because the favorable circumstances specified above have never been fully realized.
Specifically, (i) almost all known superconductors have a high enough carrier density that the magnetic field required to place a low LL index at the Fermi level is inaccessible with current magnets, and 
(ii) the vast majority of superconductors are spin singlet, and Pauli pair breaking is then strongly 
detrimental to the reentrance phenomenon.  
In principle, the latter limitation can be overcome by tilting the 
magnetic field so that the Zeeman splitting matches the Landau level separation. 
However, tilting requires even higher magnetic fields, and theory predicts that
the $T_c$ achieved is much smaller because the resulting degeneracy is between LLs with 
different orbital indices~\cite{Norman1990}.
In addition, disorder generally suppresses $T_c$ via LL 
broadening~\cite{Gruenberg1968,Norman1990} which 
lowers the enhanced density of states in partially filled LLs.
In this paper, we show that the above limitations are minimized in magic angle twisted bilayer graphene (MATBG)~\cite{Cao2018} at appropriate carrier densities.

The discovery of superconductivity and correlated insulating phases in MATBG~\cite{Cao2018,Cao2018a} has sparked interest in moir{\'e} materials as highly tunable platforms
for novel topological and correlated phases~\cite{Wu2018b,Yankowitz2019,Serlin2019,Lu2019,Chen2019,Chen2020,Liu2020,Cao2020,Lu2021,Zhang2020,Andrei2021}. 
The origin of superconductivity, whether due to electron-phonon~\cite{Wu2018,Peltonen2018,Lian2019} or electron-electron interactions~\cite{Fidrysiak2018,Kennes2018,Dodaro2018,Isobe2018,Liu2018a,You2019,Khalaf2021}, is still unsettled.  Our focus here is 
narrower and largely independent of the pairing mechanism.  We explore the 
possibility of exploiting these unusual two-dimensional superconductors, 
which have a relatively large $T_c$ at extremely low 
carrier densities, to finally achieve quantum Hall reentrant 
superconductivity.  We focus on the carrier density that corresponds to the most
robust superconducting dome observed in MATBG~\cite{Cao2018,Lu2019},
employ a simplified continuum band structure model that is  
consistent with Shubnikov-de Haas data, and combine it  
with a phenomenological BCS interaction model that is  
consistent with the observed $T_c$ and $H_{c2}$. 
We then perform LL representation particle-particle ladder sums, associating 
divergences with transitions to the 
superconducting state. In this way we obtain the full magnetic field versus temperature 
phase diagram.  The present calculations generalize previous single-band results to the qualitatively new multi-band
problem presented by MATBG.  Our findings show that at the mean-field level, reentrant quantum Hall superconductivity is achievable at moderate magnetic fields when disorder is weak. Hence, MATBG is a potential platform for the realization of a long sought phase of matter, the quantum Hall superconductor. 

\section{Model\label{Sec:model}}
The superconducting dome on which we focus lies just below the moir{\'e} band filling factor
$\nu_s=-2$~\cite{Cao2018,Lu2019}  ($\nu_s \equiv n A_M$ where $n$ is the carrier density and $A_M$ is the moir\'e 
unit cell area). 
Both experiment and theory suggest that the many-electron ground state is either valley-polarized or spin-polarized ~\cite{Saito2021,Das2021,Wu2021,Stepanov2020,Xie2020,Bultinck2020}
over the entire range of filling factor underlying the superconducting dome below $\nu_s=-2$.
The normal state Shubnikov-de Haas (SdH) data show a Landau fan emerging from $\nu_s=-2$ 
with gaps at LL filling factors $\nu_{LL} = 2,\,4,\,6,..$, 
consistent with approximate degeneracy between 
LLs distinguished by an internal index, most likely graphene valley 
(see Ref.~\cite{Prada2021} for a discussion of the 
small single-particle valley g-factors in graphene systems).

We therefore assume that the MATBG superconductors have valley-singlet pairing in a 
spin-polarized state.  This assumption is supported not only by the two-fold 
normal state LL degeneracy that survives to the highest in-plane magnetic fields, 
but also by the fact that the in-plane critical field in MATBG exceeds the perpendicular field $H_{c2\perp}$~\cite{Cao2018}, with
strong signals of spin-triplet superconductivity recently observed 
in the closely related twisted trilayer graphene~\cite{Park2021,Cao2021}
(see Appendix~\ref{Sec:spin_singlet} for a discussion of the spin-singlet case). 

\begin{figure}[h]
  \includegraphics[width=0.45\textwidth]{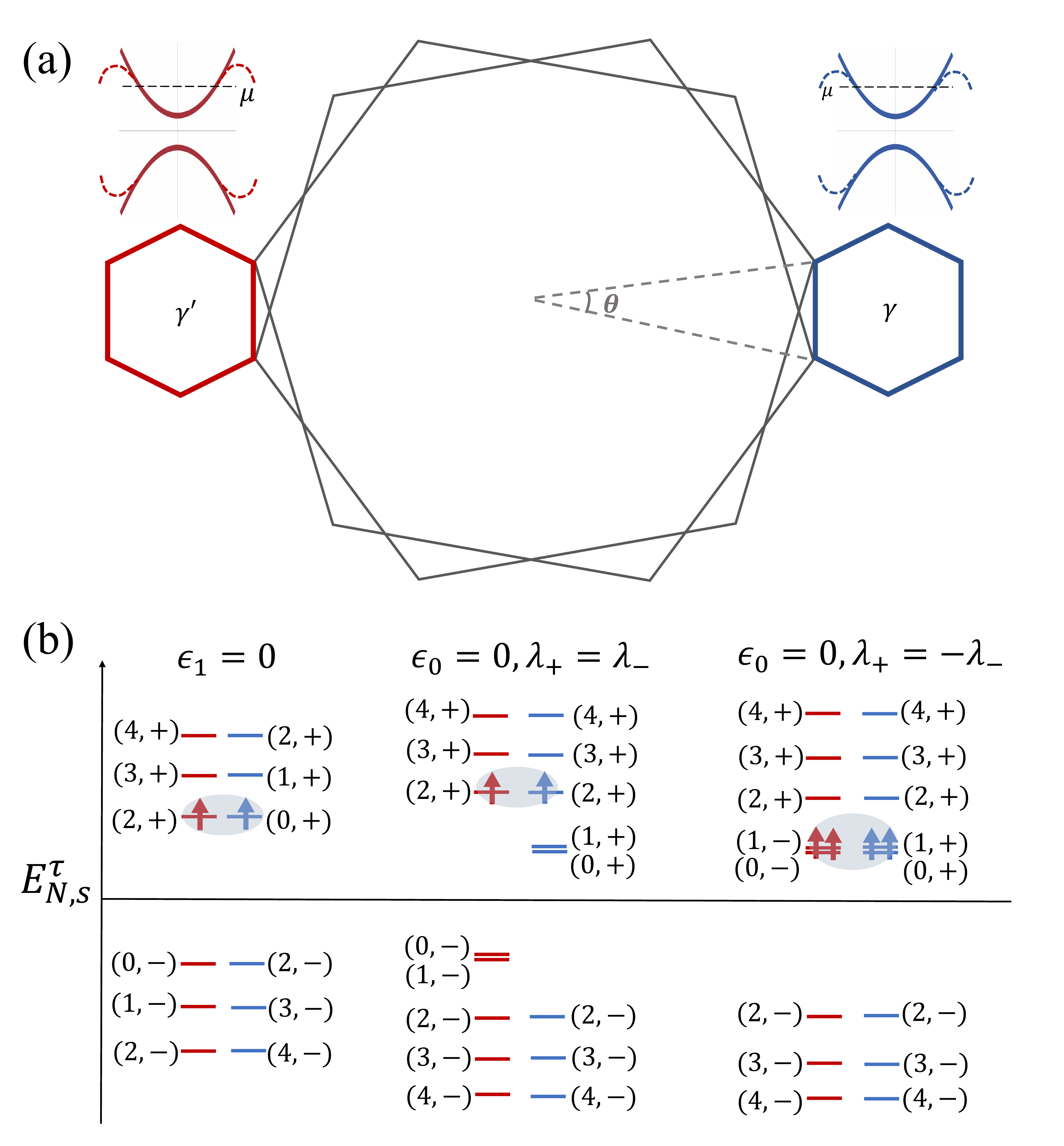}
  \caption{\label{Fig:LL_schematic}
  (a) Schematic of the band model centered around the two moir{\'e} Brillouin zone centers in opposite valleys. The dashed red and blue bands qualitatively represent the actual moir{\'e} bands, while the solid red and blue bands represent the $\vec{k}\cdot\vec{p}$ expansion employed in our model. 
  (b) Schematic quantum limit LL spectra for the three cases considered in the main text.
  For $\epsilon_1=0$, the conduction and valence band Landau levels are simply those of doubly-degenerate 
  parabolic bands. The non-standard labeling scheme employed here accounts for the anomalous $N=0,1$ Landau levels that emerge when $\epsilon_1 \ne 0$.  The blue and red levels represent `$+$' and `$-$' valley LLs, respectively.  For the massive Dirac cases ($\epsilon_0=0$), the anomalous $N=0,1$ Landau levels group with opposite bands in opposite valleys when $\lambda_{+}= \lambda_{-}$, but with the same band in both valleys when $\lambda_{+}= - \lambda_{-}$.  In all cases the LLs appear in degenerate pairs, emphasized by 
 gray ovals.  The arrows emphasize our assumption of full spin polarization.
  }
\end{figure}

Hartree-Fock band calculations~\cite{Guinea2018,Xie2020,Kang2021} suggest that when $C_{2}T$ symmetry is broken, the valence band maximum and conduction band minimum near $\nu_{s} = \pm 2$ filling occur at the moir{\'e} Brillouin zone centers, $\gamma,\, \gamma'$. 
This is further
supported by an exact diagonalization study~\cite{Potasz2021}, the observed Landau level filling factors mentioned above, as well as the absence of a Berry phase associated with quantum oscillation data~\cite{Wu2021}.  Because the quasiparticle Hamiltonian that underlies the 
superconducting state is not yet reliably known, we employ a flexible two-band two-valley band structure model that captures key qualitative features.
The model band Hamiltonian near $\gamma,\,\gamma'$ projected onto valley $\tau$ is
\begin{align}\label{Eq:k-pHam}
    & h_{\tau}(\vec{k}) = (b(\vec{k})-\lambda_{\tau})\sigma_z + \vec{d}_{\tau}(\vec{k}) \cdot \vec{\sigma}_{\parallel}\,    
\end{align}
where $b(\vec{k}) = -\hbar^2(k^2_x+k^2_y)/2m $, $\vec{d}_{\tau}(\vec{k}) = \hbar^2 v^2  (k^2_x-k^2_y,\, 2\tau k_x k_y)^{T}$, 
$\lambda_{\tau}$ is a constant that controls the band gap,
and $\sigma$ is a Pauli matrix that acts on the two-level flat-band orbital degree of freedom. 
The bands of our model are schematically shown in Fig.~\ref{Fig:LL_schematic}a. 
Superficially the model Hamiltonian resembles $AB$-stacked bilayer graphene~\cite{McCann2006}, however our model should be viewed as simply an expansion around the band minimum or maximum at the appropriate filling of the interaction dressed bands. As we will discuss later, the relevant band parameters such as the effective mass and the Fermi velocity will be determined from the experimental data.

For $m \to \infty$, $h_{\tau}(\vec{k})$ reduces to a massive quadratic Dirac model, whereas for $v \to 0$ it reduces to a parabolic band model.
Assuming full spin polarization,
this Hamiltonian describes the $\nu_s = -2$ correlated insulator, which shares 
competing topologically trivial and non-trivial insulator phases~\cite{Das2021} when its 
bands are half filled. 
At half filling, in the $m\rightarrow \infty$ limit, the $\lambda_{+} = \pm \lambda_{-}$ cases describe a valley Chern insulator ($\sigma_{xy} = 0$) and a Chern insulator ($\sigma_{xy} = 2e^2/h$), respectively.
The $v\rightarrow 0$ limit of the model captures the possibility, also apparently realized in experiment, 
that the valley-projected Hamiltonian is trivial on its own.
Our conclusions depend mainly on the carrier density and zero field topology; their sensitivity to 
band dispersion details that are not accurately rendered is discussed later. 
In particular, our model is isotropic and particle-hole symmetric as a matter of convenience.  
We include only the two partially-filled bands in evaluating the ladder sums, with filled and empty bands
considered as inert.  Neither approximation should cause any qualitative changes in our results.

Reentrant superconductivity emerges in mean-field theory when an enhanced Fermi level density of states, 
produced by Landau quantization, trumps broken time-reversal symmetry.
It is therefore captured only by a fully quantum treatment of the perpendicular magnetic 
field $B\hat{z}$.
Using a Landau gauge for the vector potential $\vec{A} = (-By,\,0,\,0)$, the band 
Hamiltonian becomes 
\begin{align}\label{Eq:magneticHam}
    & h_+(\hat{a},\hat{a}^{\dagger}) = \begin{pmatrix}
    -\epsilon_0(\hat{a}^{\dagger}\hat{a} + \frac{1}{2}) - \lambda_+ & \epsilon_1 (\hat{a})^2\\
     \epsilon_1 (\hat{a}^{\dagger})^2 & \epsilon_0(\hat{a}^{\dagger}\hat{a}+\frac{1}{2})  + \lambda_+
    \end{pmatrix} \, ,
\end{align}
where $\epsilon_0 = \hbar^2/(m\ell^2)$, $\epsilon_1 = 2\hbar^2 v^2/\ell^2$, and $\ell = \sqrt{\hbar/e B}$ is the magnetic length. Here $\hat{a} = (\hat{K}_x- i\hat{K}_y) \ell/\sqrt{2}$ is the LL lowering operator, 
$\hat{\vec{K}} = \vec{k}+e\vec{A}$,
and $e$ is the magnitude of the electronic charge ($h_-$ for the $-$ valley is obtained by the replacement $a \to - a^{\dagger}$ in the off-diagonal terms).
After diagonalization of the single-particle Hamiltonian, we obtain the LL spectrum
\begin{align}\label{Eq:LL_spectrum}
    E^{\tau}_{N,s} &= \begin{cases} 
    \tau\epsilon_0 +
    s \biggl (\epsilon^2_1 N (N-1) \\
    \hspace{1.3cm} + [\epsilon_0(N-\frac{1}{2})+\lambda_{\tau}]^2\biggr )^{1/2}
     \quad\quad \text{$N \geq 2$}\\
      \tau\epsilon_0 (N+\frac{1}{2}) + \tau\lambda_{\tau}\, ,  \quad\quad \text{$s = \tau$\, and  $N = 0,\, 1$}
    \end{cases} \, 
\end{align}
where $s = \pm $ distinguishes conduction (higher energy) and valence (lower energy) bands. 
A schematic of the LL spectrum is presented in Fig.~\ref{Fig:LL_schematic}b. 
Note that the anomalous $N = 0,\, 1$ LLs can be energetically close to
the conduction or valence band levels, depending on the model.  
For example, in the $\epsilon_0\rightarrow 0$ limit, when $\lambda_+ = \lambda_-$ these LLs group with the upper (lower) band in valley $+$ ($-$), and when $\lambda_+ = -\lambda_-$ these LLs group with the upper band in both valleys.
For the sake of brevity, we define $\vec{N} = (N,s)$. 
 The band LL eigenfunctions are~\footnote{For the $\epsilon_1 = 0$ case, the expression for $\alpha^{\tau}_{\vec{N}}$ can have vanishing numerator and denominator. In this case, $\alpha^{\tau}_{\vec{N}} = 0$.}
\begin{subequations}\label{Eq:LL_states}
\begin{align}
   & \psi^{+}_{\vec{N},Y}(\vec{r}) =  \biggr (\alpha^{+}_{\vec{N}} \phi_{N-2,Y}(\vec{r}),\, \beta^{+}_{\vec{N}}\phi_{N,Y}(\vec{r})   \biggr )^T\, , \\
   & \psi^{-}_{\vec{N},Y}(\vec{r}) =  \biggr (\beta^{-}_{\vec{N}} \phi_{N,Y}(\vec{r}),\, \alpha^{-}_{\vec{N}}\phi_{N-2,Y}(\vec{r})   \biggr )^T\,
\end{align}
\end{subequations}
where 
\begin{align}\label{Eq:LL_coeff}
    \alpha^{\tau}_{\vec{N}} = \begin{cases}
    0 &\text{$ N = 0,\,1 $}\\
    \frac{s(E^{\tau}_{\vec{N}}-\tau\epsilon_0(N+1/2)-\tau\lambda_{\tau} )}{\sqrt{(E^{\tau}_{\vec{N}}-\tau\epsilon_0(N+1/2)-\tau \lambda_{\tau} )^2+\epsilon^2_1N(N-1)}} & \text{$N\geq 2$ }\, ,
    \end{cases}
\end{align}
$\beta^{\tau}_{\vec{N}} = \sqrt{1-|\alpha^{\tau}_{\vec{N}}|^2}$, and $\phi_{n,Y}(\vec{r}) = \textup{e}^{ixY/\ell^2}\,
\varphi_{n}(y/\ell-Y/\ell)/\sqrt{L_x}$ is a parabolic band $n^{th}$ LL wavefunction 
with $\varphi_n$ a 1D harmonic oscillator wavefunction and $Y$ the LL guiding center.

\section{Cooper Instability\label{Sec:Cooper}}
We are interested in Cooper instabilities when Landau quantization of the pairing 
electrons is fully taken into account. 
For simplicity, we choose a model with an attractive interaction $\hat{V}$
between electrons in different valleys.  
We set $\hat{V}  = -V\delta(\vec{r})$, implying an interaction range that is short compared to the 
Fermi wavelength, but long compared to the microscopic graphene lattice constant and the valley-singlet superconducting coherence length. We focus on the limits in which electrons interact only when they are in the same flat band orbital, or only in opposite flat band orbitals.
The linearized $T_c$ equations derived from the particle-particle ladder sum (see Appendix~\ref{Sec:Ladder} for details) are
\begin{align}\label{Eq:instability_main}
    & X_j = 1\, \quad\quad\quad\text{(intra-orbital interactions)}\, ,\notag\\
    & Z_j = 1\,\quad\quad\quad \text{(inter-orbital interactions)}\,
\end{align} 
where $X_j$ and $Z_j$ are given by  
\begin{subequations}\label{Eq:linearizedTc}
\begin{align}
    & X_j = -\frac{\lambda_0}{4\pi\ell^2\rho}\sum_{\tau,\vec{N},\vec{M}}  \mathcal{K}^{\tau}_{\vec{N},\vec{M}}(i\omega\rightarrow 0)\alpha^{\tau}_{\vec{N}}\beta^{-\tau}_{\vec{M}}\mathcal{B}^{N-2,M}_j\notag\\
    &\hspace{1cm}\times[\alpha^{\tau}_{\vec{N}}\beta^{-\tau}_{\vec{M}}\mathcal{B}^{N-2,M}_j + \beta^{\tau}_{\vec{N}}\alpha^{-\tau}_{\vec{M}}\mathcal{B}^{N,M-2}_j]\, , \label{Eq:linearizedTc1}\\
    & Z_j = -\frac{\lambda_0}{8\pi\ell^2\rho}\sum_{\tau,\vec{N},\vec{M}}  \mathcal{K}^{\tau}_{\vec{N},\vec{M}}(i\omega\rightarrow 0)[(\beta^{\tau}_{\vec{N}}\beta^{-\tau}_{\vec{M}}\mathcal{B}^{N,M}_j)^2\notag\\
    &\hspace{3cm}+ (\alpha^{\tau}_{\vec{N}}\alpha^{-\tau}_{\vec{M}}\mathcal{B}^{N-2,M-2}_j)^2]\, \label{Eq:linearizedTc2}.
\end{align}
\end{subequations}
Here, $\lambda_0>0$ is the BCS coupling constant, $\rho$ is the zero field density of states at the Fermi level, and the coefficients 
\begin{align}\label{Eq:transCoeff}
    & \mathcal{B}^{N,M}_j = \sum^j_{m=0} (-)^{M-m}\sqrt{\frac{{^j}C_m {^M}C_m {^N}C_{j-m} {^{N+M-j}}C_{M-m}}{2^{N+M}}}  \,  ,
\end{align}
where ${^n}C_k = n!/[k!(n-k)!]$ arise from the unitary transformation~\cite{MacDonald1992} between the two-particle product state and the center-of-mass (C.O.M.) and relative LL wavefunctions.
In the equations for $X_j$ and $Z_j$,
\begin{align}\label{Eq:Kern}
     & \mathcal{K}^{\tau}_{\vec{N},\vec{M}}(i\omega) = \int\int d\epsilon_1 d\epsilon_2 \frac{\tanh(\beta \epsilon_1/2)+\tanh(\beta \epsilon_2/2)}{2(i\omega-\epsilon_1-\epsilon_2)} \notag\\
     &\hspace{1cm}\times A(\epsilon_1-(E^{\tau}_{\vec{N}}-\mu))A(\epsilon_2-(E^{-\tau}_{\vec{M}}-\mu))
\end{align}
is the pair propagation kernel, $\mu$ is the chemical potential, $\beta = 1/(k_B T)$,
and $A(\epsilon)$ is the disorder-broadened LL spectral function.

Our derivation is an extension of Ref.~\cite{MacDonald1992} to the more 
general two-band case.
The most important observation is that the ladder sum separates into channels labeled by  the C.O.M. LL index $j$.
Typically, the $j = 0$ channel yields the highest $T_c$~\cite{Akera1991}.

In Eq.~\ref{Eq:Kern}, we have replaced the $\delta$-function spectrum of each unbroadened Landau level by a Lorentzian $A$.  Broadening smears the $1/T$ divergence 
that appears in the pairing kernel at fractional LL filling factors when 
opposite valley LLs are degenerate. In principle, broadening can arise either from disorder or from the periodic 
moir\'e potential.  The total effective Lorentzian broadening factor, $\eta$, is fixed 
by experimental Shubnikov-de Haas data using $\eta=\pi k_B T_D$  where $T_D$ is the 
Dingle temperature (we ignore other pair breaking effects related to disorder, 
since they are absorbed when we set model parameters to match the zero-field $T_c$).
In evaluating Eq.~\ref{Eq:Kern}, we assume a pairing window $\pm\omega_D$ around the chemical potential, and choose the value of $\omega_D$ to match the low-field experimental data. 
To avoid spurious effects due to a hard cut-off, we introduce Gaussian weighting factors for each LL:
$W_\vec{N}=\sqrt{1.55}e^{-((E_\vec{N}-\mu)/\omega_D)^2}$, with $W_\vec{N}W_\vec{M}$ entering as a 
prefactor in Eq.~\ref{Eq:Kern}.

\section{Reentrant superconductivity in MATBG\label{Sec:ReentrantSC}}
In MATBG there are in total eight flat bands, counting spin, valley and flat-band orbital degrees of freedom - four conduction bands and four valence bands.  It is generally agreed that 
the MATBG insulators which appear at integer filling factors can be viewed as 
Slater insulators in which bands are either fully occupied or empty~\cite{Xie2020} and occupancies are 
spin/valley flavor dependent.  Hence at $\nu_s = -2$ two of the eight flat bands are occupied. 
Most experiments suggest that the occupied bands are doubly degenerate~\cite{Cao2018,Uri2020,Wu2021} (although exceptions exist~\cite{Das2021}) and therefore that the ground state is likely to be ferromagnetic.
For the MATBG superconducting state at the moir{\'e} filling fraction $\nu_s = -2-\delta $, 
we study the three cases illustrated in Fig.~\ref{Fig:LL_schematic}b.

\smallskip\noindent
i) $v =  0$: 
In this case the insulating state is topologically trivial. 
We can replace $m$ in $b(\vec{k})$ by the effective electron cyclotron mass 
$m^{\ast}$. We estimate $m^{\ast}$ by relating it to $T_c$ and the 
semiclassical $H_{c2\perp}$ using~\cite{Gorkov1960} 
\begin{equation}
    m^{\ast} = \frac{\hbar}{k_B T_c}\sqrt{\frac{\pi e n_{2D}H_{c2\perp}}{5.54}}
\label{eq:mstar}
\end{equation}
where $n_{2D}\sim 2.8\times 10^{11} \textup{cm}^{-2}$ is the carrier density measured from the $\nu_s = -2$ correlated insulator.
Using $H_{c2\perp}\sim 0.3 T$ and $T_c\sim 3K$~\cite{Lu2019}
yields $m^{\ast} \sim 0.27 m_e$, which is in good agreement with the effective mass 
extracted from Shubnikov-de Haas oscillations by Cao \textit{et al.}~\cite{Cao2018}.
To match the zero field $T_c\sim 3K$ and the 
semiclassical $H_{c2\perp}\sim 0.3 T$, 
we choose $\omega_D \sim 30 K$ and a BCS coupling constant $\lambda_0 \sim 0.4 $. 
In this limit the LLs with the same energy have the same orbital index in opposite valleys. 
Hence only intra-orbital interactions are relevant.

\smallskip\noindent 
ii) $m\rightarrow \infty, \lambda_+ = \lambda_-$: In this case the insulating state is a valley Chern insulator with Hall conductivity $\sigma_{xy} = 0$.
We obtain the same $B=0$ dispersion  in the gapless limit as in case i 
by setting $v = 1/\sqrt{2m^{\ast}}$. We then add a mass gap $\lambda_{\pm} \sim 25K$ and let the other model 
parameters retain the same values as in case i. As before, we consider intra-orbital interactions.

\smallskip\noindent 
iii) $m\rightarrow \infty, \lambda_+ = -\lambda_-$: In this case the insulating state in our band model (Eq.~\ref{Eq:k-pHam}) is 
a Chern insulator with Hall conductivity $\sigma_{xy} = \pm 2e^2/h$.
We keep all the parameters in the model to be the same as in ii), except that
$\lambda$ takes opposite signs in the two valleys. In this case we get results close to
those of the valley Chern insulator case by considering only inter-orbital interactions (with an appropriately defined BCS coupling constant, twice that of the intra-orbital one). 
For intra-orbital interactions, we get a result similar to the valley Chern case.

\begin{figure}[h]
  \includegraphics[width=0.45\textwidth]{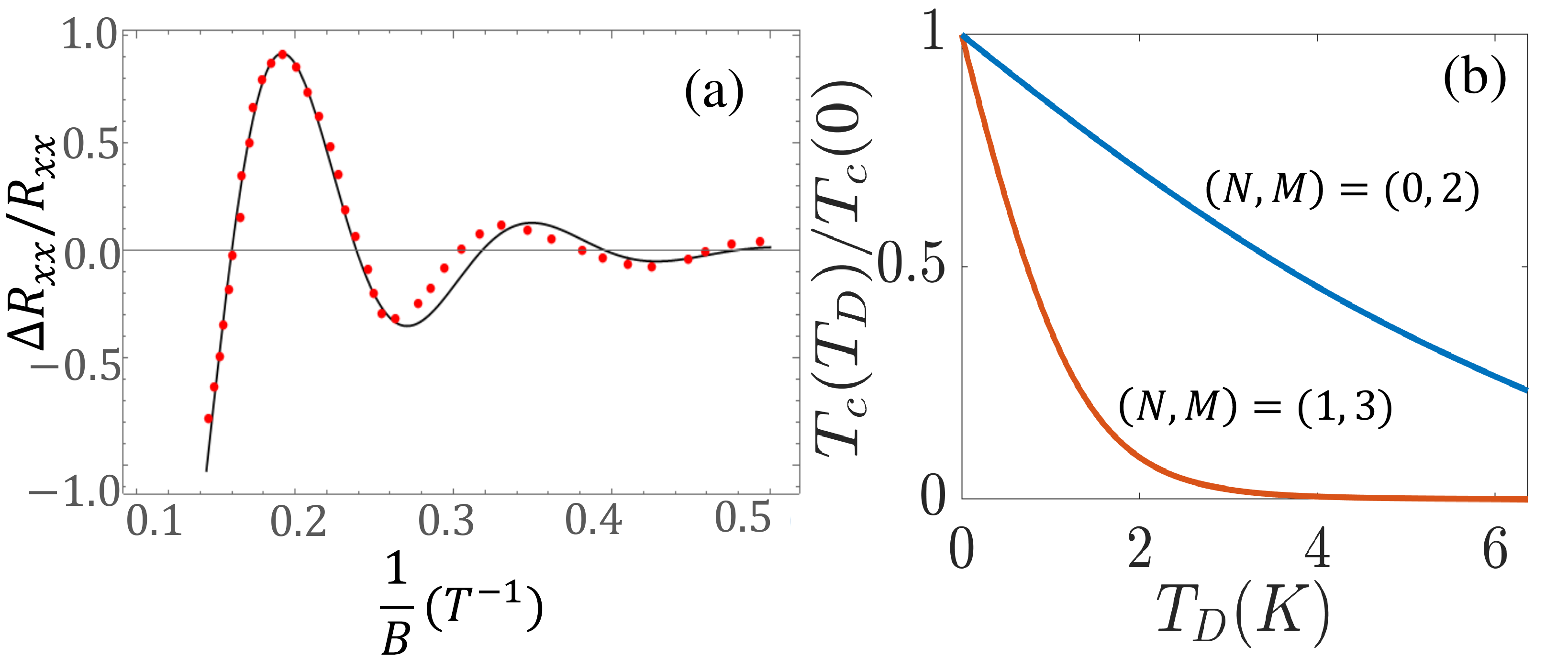}
  \caption{\label{Fig:Broadening_spin}
  (a) Dingle temperature estimate for MATBG. The red dots are extracted from Shubnikov-de Haas data presented in Fig.~5b of Ref.~\cite{Cao2018} and the black curve is a Lifshitz-Kosevich fit. 
  Based on the fit, a Dingle temperature $T_D\sim 2.6 K$ is estimated. 
  (b) Dependence of $T_c$ on LL broadening
  (characterized by the Dingle temperature) in the reentrant regime for the 
  parabolic band (i.e. $\epsilon_1 = 0$) limit of our model. 
 }
\end{figure}

Before discussing our results, we comment on the crucial Dingle temperature estimates. 
In Fig.~\ref{Fig:Broadening_spin}a, we plot the Shubnikov-de Haas data at $T = 0.7K$, obtained from Ref.~\cite{Cao2018} and fit to a Lifshitz-Kosevich expression supplemented by a Dingle factor
\begin{align}
    & \frac{\Delta R}{R}\propto \sin \biggl (\frac{2\pi F}{B} + \theta_0 \biggr ) \frac{c T/B}{\sinh (c T/B)} \textup{e}^{-c T_D/B}\, ,
\end{align}
where $ c = 2\pi^2m^{\ast}k_B/(e\hbar) $.
For the data shown in Fig.~\ref{Fig:Broadening_spin}a, we estimate $T_D\sim 2.6 K$.
The resultant LL broadening in Eq.~\ref{Eq:Kern} is $\eta = \pi k_B T_D $.
We mention that the samples in Ref.~\cite{Liu2020} are likely
cleaner and thus should exhibit a smaller Dingle temperature.
Unfortunately, Ref.~\cite{Liu2020} does not report Shubnikov-de Haas data.

\begin{figure*}[t]
  \includegraphics[width=1\textwidth]{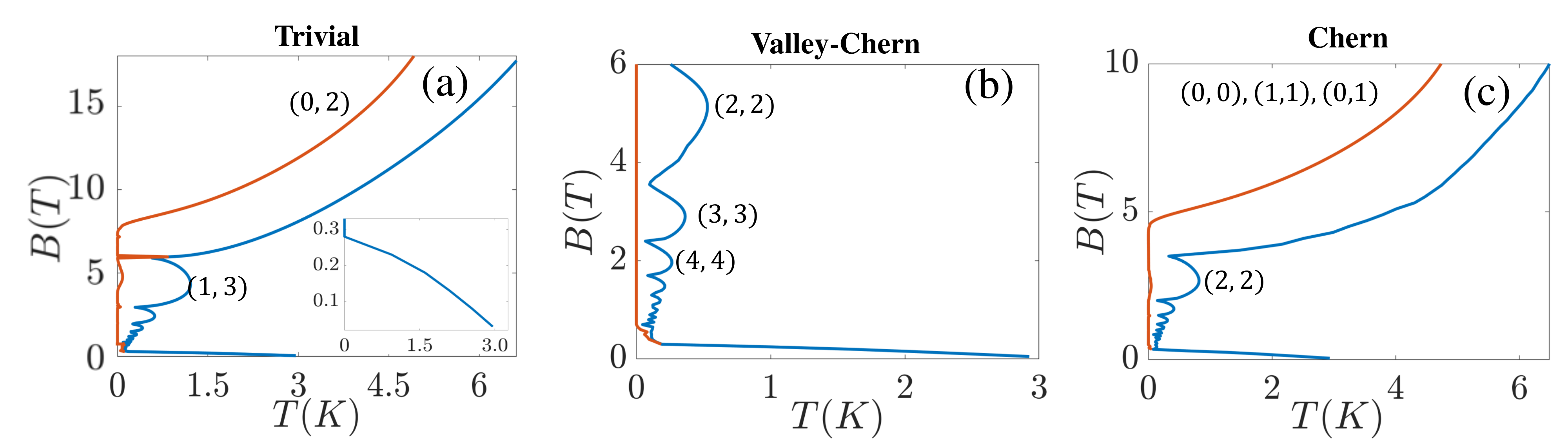}
  \caption{\label{Fig:BTPhase}
  Magnetic field versus temperature phase diagram at fixed density for the three cases
  illustrated in Fig.~1b: (a) $v=0$, (b) $m\rightarrow\infty,\, \lambda_+ = \lambda_-$, and (c) $m\rightarrow\infty,\,\lambda_+ = -\lambda_-$.
  The blue lines represent the disorder-free limit, and the red lines take disorder broadening into account using a Dingle temperature $T_D = 2.6 K$.
  The inset in (a)  expands the low field region near the semiclassical $H_{c2}$.
  The plots show robust reentrant superconductivity when the quantum limit is reached at moderate magnetic fields 
  for case i (trivial) and case iii (Chern).
  The most robust domes at $\gtrsim 10T$ fields correspond to pairing in the lowest LLs.
  The qualitative difference between (b) and (c) is due to the absence in the former case (ii) 
  of degenerate intervalley $N=0, 1$ pairs.}
\end{figure*}

In Fig.~\ref{Fig:BTPhase} we show magnetic field versus temperature phase diagrams calculated by solving Eqs.~\ref{Eq:linearizedTc} and \ref{Eq:Kern} at fixed electron density using the model parameters discussed above.
The chemical potential $\mu(T=T_c)$ oscillates as a function of field to fix the carrier density at $n_{2D}$, 
and at low temperature tends to get pinned at one of the Landau level energies.
In all cases, we see clear reentrant superconductivity once the magnetic field becomes 
high enough that the single pair of LLs closest to the chemical potential dominates the pairing. 
For case i (with intra-orbital interactions) and the `Chern' case iii (with inter-orbital interactions),
the reentrant phase appears at relatively moderate magnetic fields
when the LL broadening is set by the 
experimental Dingle temperature. 
However for case ii (with intra-orbital interactions), in which the zero field model is a valley Chern insulator at
$\nu_s =-2$, the reentrant phase is more fragile to disorder. 
We explain this difference below.

Quantum Hall reentrant superconductivity is strongest when a degenerate pair of LLs
distinguished by valley is each half filled.
The LL index at the Fermi level decreases and the LL degeneracy increases with increasing magnetic field.
Both cause the reentrant phase to become more robust as the field increases. 
These trends are easiest to understand 
in the extreme high field limit in which a single pair of LLs dominates
the pairing kernel. 
In
the half-filled topologically trivial case (Fig.~\ref{Fig:LL_schematic}b left), 
LL $N$ in the upper band of the `$+$' valley is degenerate with LL $N+2$ in the upper band of the `$-$' valley and pair when the interactions are attractive.
In the half-filled topologically non-trivial normal state case, 
LLs $N\geq 2$ in the two valleys are degenerate ($N = 0,\,1 $ will be discussed below). Hence the pair is formed between LL $N$ in both valleys. Assuming that the LLs are close enough to the chemical potential that we can 
replace $\tanh(x)$ by $x$ in the kernel, we can further simplify Eq.~\ref{Eq:linearizedTc1} and  Eq.~\ref{Eq:Kern}.  For intra-orbital interactions, we obtain
\begin{align}
     & k_B T_c = \frac{\lambda_0}{16 \pi \ell^2 \rho} 
     c_{N,j} (\mathcal{B}^{N-\delta,N}_j)^2\, 
\end{align}
where $\delta = 0$ for the $\epsilon_1 = 0$ (trivial) case and 
$\delta = 2$ for the $\epsilon_0 = 0$ (non-trivial) case.
The proportionality constant $c_{N,j}$ depends on the band model and LL diagonalization coefficients $\alpha^{\tau}_\vec{N},\beta^{\tau}_\vec{N}$.  For example, in the trivial case, $c_{N,j} = 1$, and in the non-trivial case with intra-orbital interactions, $c_{N,j} =  [ \alpha^{+}_{\vec{N}}\beta^{-}_{\vec{N}}+ (-1)^j \alpha^{-}_{\vec{N}}\beta^{+}_{\vec{N}}]^2$. 
Since $(\mathcal{B}^{N-\delta,N}_0)^2 = (\mathcal{B}^{N-\delta,N}_{2N-\delta})^2$, we get two C.O.M. channels $j = 0,\, 2N-\delta$ with equal $T_c$. 
The solution to the non-linear gap equations is a vortex lattice~\cite{Akera1991}. 
For $j$ non-zero, the vortex lattice differs qualitatively from the $j=0$ Abrikosov solution~\cite{Akera1991}.
 As shown in Fig.~\ref{Fig:BTPhase} and discussed below, we obtain robust quantum Hall superconductivity in the lowest two LLs. It follows that the $j=0$ solution is the relevant one, as we will assume from this point forward.
 
 
We see from Eq.~12 that $T_c$ has an explicit linear dependence on magnetic field and an additional implicit dependence via the transformation coefficient $\mathcal{B}^{N,M}_0$, which takes larger values at smaller LL index since
$(\mathcal{B}^{N,M}_0)^2={^{N+M}}C_{N}/2^{N+M}$.
The dependence of $T_c$ on the zero-field BCS coupling constant is linear 
because of the $1/T$ divergence of the pairing kernel.
The reentrant $T_c$ can therefore
even exceed the zero field $T_c$, at least in the absence of LL broadening.

In the half-filled valley Chern insulator  
case (Fig.~\ref{Fig:LL_schematic}b middle), the $N = 0,\, 1$ LLs are degenerate in each valley but separated by an 
energy 2$\lambda$. Hence, in this case carriers become valley polarized 
at strong fields and cannot form intervalley pairs. 
In contrast, for the half-filled Chern insulator case (Fig.~\ref{Fig:LL_schematic}b right), the $N = 0,\, 1$ LLs in opposite valleys are degenerate again. However, the $N=0,\, 1$ LLs of the opposite valley have opposite orbital polarization. 
Hence, attractive inter-orbital interactions are needed to support robust reentrant superconductivity (Fig.~\ref{Fig:BTPhase}c).

In Fig.~\ref{Fig:Broadening_spin}b, we illustrate $T_c$ suppression by LL broadening for the
trivial case in the approximation that a single pair of LLs saturates the pairing kernel. Although $T_c$ is suppressed quite rapidly for the second lowest LL (requiring very clean samples for its observation), the lowest LL $T_c$ is much more robust.
Since in case ii (Chern) the lowest two LLs cannot host pairing, 
and the reentrant $T_c$ for the other levels is small,  reentrant quantum Hall superconductivity for this case becomes fragile to disorder.

\section{Discussion\label{Sec:Discussion}}
We have employed a simplified model band structure that is motivated by the observation in experiment 
of a simple pattern of Shubnikov-de Haas oscillations indicative of two degenerate closed hole-like 
Fermi pockets.  Our model is flexible enough to allow for underlying bands with different types of 
topological character, and we analyzed the three different possibilities - bands with trivial topology, valley Hall bands,
and Chern bands.  The model certainly fails at very 
strong magnetic fields where $1/\ell$ becomes comparable to the moir\'e Brillouin zone dimensions.
Strong field modifications in the LL structure are visible in the Shubnikov-de Haas data, 
where quantum oscillations disappear once the Fermi surface approaches the van Hove ($M$) points of the zone~\cite{Cao2018}. 
Although Fig.~\ref{Fig:BTPhase} gives the impression that $T_c$ indefinitely increases with $B$ (due to increasing level degeneracy), in reality $T_c$ has an ultraviolet cut-off from the flat band width.  

Although we have shown that MATBG could be an ideal platform for observing reentrant quantum Hall superconductivity,
magneto-transport data reported in these materials have not to date 
given any evidence of reentrant superconductivity~\cite{Cao2018,Liu2020}.
The reported data are typically below $7-8T$, and so the quantum limit 
has not been reached.  Optimistically, it is possible that simply going to a larger field $B\sim 10-15T$
will allow the reentrant state to emerge.  
This conclusion is far from certain, however, for a number of reasons.
For example, as discussed above, superconductivity in the quantum Hall limit 
is suppressed by LL broadening which can originate from disorder and/or a moir{\'e} potential.
We have estimated the Dingle temperature  based on available data~\cite{Cao2018} (which in principle subsumes both effects) and find only a moderate suppression of $T_c$ when pairing occurs in the lowest LL.
We expect that even smaller Dingle temperatures have been achieved 
in more recent work (Ref.~\cite{Lu2019}), implying an even smaller suppression of $T_c$.
On this basis we judge that disorder does not stand in the way of this long-sought phenomenon.  
However, more fundamental obstacles could prove important.

Our calculations predict reentrant quantum Hall superconductivity in MATBG provided that 
the factors that justify the BCS mean-field theory at $B=0$ are not undone by the magnetic field.
Analyzing the influence of quantum and thermal fluctuations on quantum Hall superconductivity is challenging~\cite{Rasolt1992}.  
It is thought that LL quantization reduces the effective dimensionality of the system, which in turn 
suppresses superconductivity due to fluctuations (long range order does not exist in dimensions below two).
Thermal fluctuations primarily act to melt the vortex lattice.  
A deleterious influence of thermal fluctuations might be 
ameliorated by the moir{\'e} periodic potential which will pin the vortex lattice and act
to restore the dimensionality - although it also suppresses $T_c$ because of LL broadening.
Moreover, the topological nature of the LLs might also give an additional contribution to the superfluid density~\cite{Peotta2015,Liang2017,Wang2020a}.
Hence, the role of fluctuations in suppressing the long range order might be smaller than expected.  
Regardless, transport signatures of fluctuating superconductivity and vortex liquids are well understood, including their nonlinear $I-V$ characteristics, and thus potentially observable even if long range order does not occur.
A notable example is recent work on cuprates that finds fragile superconductivity persisting to very high fields as long as low measurement currents are employed~\cite{Hsu2021}. 
Moreover, spectroscopic signatures of a gap would likely be present as well, as seen above $T_c$ in cuprates, as well as in
materials near a superconductor-insulator transition~\cite{Sacepe2020}.
Another concern is that the reentrant phase could be unstable to other correlated phases that might be 
stabilized by higher fields.  Moreover, the pairing problem in this quantum limit is in reality non-adiabatic in nature, in that the only relevant energy scale is the separation of the chemical potential from the LL energy, so one naturally expects 
deviations from mean-field theory that are challenging to describe theoretically and beyond the scope of this paper.

In our analysis, the effect of the moir{\'e} potential is taken into account via a simple LL broadening.
On the other hand, when our predicted quantum Hall superconductivity regime sets in, the magnetic length  $\ell$ and the moir{\'e} period $a_M$ start to become comparable, so the low lying LLs experience the moir{\'e} potential, forming
a Hofstadter butterfly spectrum at sufficiently high magnetic fields.
These modified LLs should not only affect the equation for $T_c$, they will affect the non-linear equations and thus potentially the form of the vortex lattice, whose length scale is controlled by $\ell$.
This makes moir{\'e} quantum Hall superconductivity even richer~\cite{Shaffer2021}, with no precedent in previous studies. In particular, we expect that if quantum Hall superconductivity is achieved, tuning of $a_M/\ell$ could lead to further topological phase transitions. A detailed study of these non-linear effects are beyond the scope of this work and is left for future investigations.

In this work, we focused on MATBG because the phenomenology of its superconductivity has been thoroughly studied.
However, many essential features are expected to be replicated in other moir{\'e} systems, for example in
superconducting twisted trilayer graphene~\cite{Park2021}.
We expect that as this field progresses, there may appear other moir{\'e} materials 
that offer even better conditions for observing the reentrant phase.

In conclusion, we have shown that moir{\'e} superconductors may be an ideal platform to achieve reentrant superconductivity in the quantum limit.
In the future, a rigorous treatment of fluctuations in the high field limit would be desirable, along with a related estimate of transport signatures.
Understanding the competition between the repulsive Coulomb and the attractive pairing interactions in the lowest LL limit is also an interesting direction to pursue, the former leading to the fractional quantum Hall effect, the latter to superconductivity.
An interplay between the two may pave the way for exotic parafermions modes~\cite{Alicea2016,Guel2020} and topological quantum computation. 
More generally, we have presented a two-band formulation for pairing in LLs, which can be an important tool to study high magnetic field effects for other two-dimensional superconductors.

\textbf{\textit{Acknowledgments}}. Work at Argonne is supported by the US DOE, Office of Science, Basic Energy Sciences, Materials Sciences and Engineering Division.  AHM was supported by the Army Research Office under Grant Number W911NF-16-1-0472.

\begin{widetext}

\appendix

\section{Evaluation of the ladder sum\label{Sec:Ladder}}
In this Appendix, we evaluate the particle-particle ladder sum for a two-band model. 
In the main text, our single-particle Hamiltonian is quadratic in momentum. 
At a different filling fraction, such as near $\nu_s = 1$, the LL spectrum suggests a massive linear-Dirac spectrum instead.
In principle, the superconducting states appearing near different fillings can also lead to reentrant quantum Hall superconductivity.
Motivated by this, we start with a more general form of an effective Hamiltonian projected in valley $\tau$ 
\begin{align}\label{Eq:Supp_kp}
    & h_{\tau}(\vec{k}) = \begin{pmatrix}
    \frac{\tilde{s}\hbar^2}{2m} (k^2_x+k^2_y) - \lambda_{\tau} & [\hbar v(k_x - ik_y)]^{\gamma}\\
    [\hbar v (k_x + i k_y)]^{\gamma} & \frac{\hbar^2}{2m}(k^2_x+k^2_y)+\lambda_{\tau}
    \end{pmatrix}\, .
\end{align}
Here $\tilde{s} = \pm$.
In the main text, we considered the $\gamma = 2$ and $\tilde{s} = -$ case. 
Here, for generality, we allow $\gamma = 1,\, 2$ and do not fix the value of $\tilde{s}$. As shown schematically in Fig.~\ref{Fig:Band_schematic}, the various parameters of the Hamiltonian can be tuned to mimic different band structures. 
Hence our derivation below can be used for a full quantum treatment of magnetic field in two band superconductors with appropriate choice of band parameters. 
After including the magnetic field using the Landau gauge, we obtain the valley $+$ Hamiltonian
\begin{align}\label{Eq:supp_magneticHam}
     & h_{+} (a,a^{\dagger}) = \begin{pmatrix}
     \tilde{s}\epsilon_0(a^{\dagger}a +\frac{1}{2}) - \lambda_{\tau} & \epsilon_1 (a)^\gamma \\
     \epsilon_1 (a^{\dagger})^{\gamma} & \epsilon_0(a^{\dagger}a+\frac{1}{2}) + \lambda_{\tau}
     \end{pmatrix}\, .
\end{align} 
The LL eigenstates in valley `$+$' are
\begin{subequations}\label{Eq:Supp_LL_states}
\begin{align}
   & \psi^{+}_{\vec{N},Y}(\vec{r}) =  \begin{cases}\biggr (\alpha^{+}_{\vec{N}} \phi_{N-\gamma,Y}(\vec{r}),\, \beta^{+}_{\vec{N}}\phi_{N,Y}(\vec{r})   \biggr )^T & \text{$N\geq \gamma $} \, ,\\
   \biggl (0, \phi_{N,Y}(\vec{r}) \biggr )^T\,  \text{$\gamma > N \geq 0$}.
   \end{cases}
\end{align}
\end{subequations}
The diagonalization coefficients $\alpha^{\tau}_{\vec{N}}$ and $\beta^{\tau}_{\vec{N}}$ for the specific case of $\gamma = 2$ and $\tilde{s} = -$ are stated in Eq.~\ref{Eq:LL_coeff} of the main text.
Our derivation of the particle-particle ladder sum below shows that for the Cooper instability, all the band dependent properties are absorbed in some simple combination of these LL diagonalization coefficients. 

To construct the two-particle states, we consider the non-interacting two-particle Hamiltonian
\begin{align}\label{Eq:Supp_Ham_twoBody}
    & H = \sum_{\tau} h_{\tau}(\hat{a}_1,\hat{a}^{\dagger}_1) + h_{-\tau}(\hat{a}_2,\hat{a}^{\dagger}_2)\, ,
\end{align}
which can equivalently be represented in the center of mass (C.O.M.) and relative coordinate basis after the transformation
\begin{align}\label{Eq:Supp_Ham_comRelative}
     & \hat{a}_R = \frac{\hat{a}_1+\hat{a}_2}{\sqrt{2}},\quad 
     \hat{a}_r = \frac{\hat{a}_1-\hat{a}_2}{\sqrt{2}}\, .
\end{align}
This transformation is made because the underlying vortex lattice solution involves only the C.O.M. degrees of freedom, whereas the pairing interaction depends only on the relative coordinates.
The transformation relation stated in terms of harmonic LL wavefunctions is
\begin{align}\label{Eq:Supp_Transformation}
    & \phi_{N,Y_c+Y_r/2}(\vec{r}_1) \phi_{M,Y_c-Y_r/2}(\vec{r}_2)
    = \sum^{N+M}_{j=0} \mathcal{B}^{N,M}_j  \phi^{R}_{j,Y_c}(\vec{R}) \phi^r_{N+M-j,Y_r}(\vec{r})\, ,
\end{align}
where $\vec{R} = (\vec{r_1}+\vec{r_2})/2$, $\vec{r} = \vec{r_1}-\vec{r_2}$, and $j$ denotes the C.O.M. LL index.
$\phi^{R}$ and $\phi^{r}$ are identical to the individual electron LL wavefunctions, once the change in the respective effective magnetic length $\ell^{R} = \ell/\sqrt{2}$, $\ell^r = \sqrt{2}\ell$ is taken into account. The transformation coefficients $\mathcal{B}^{N,M}_j$ are stated in Eq.~\ref{Eq:transCoeff} of the main text.

\begin{figure}[!t]
  \includegraphics[width=1\textwidth]{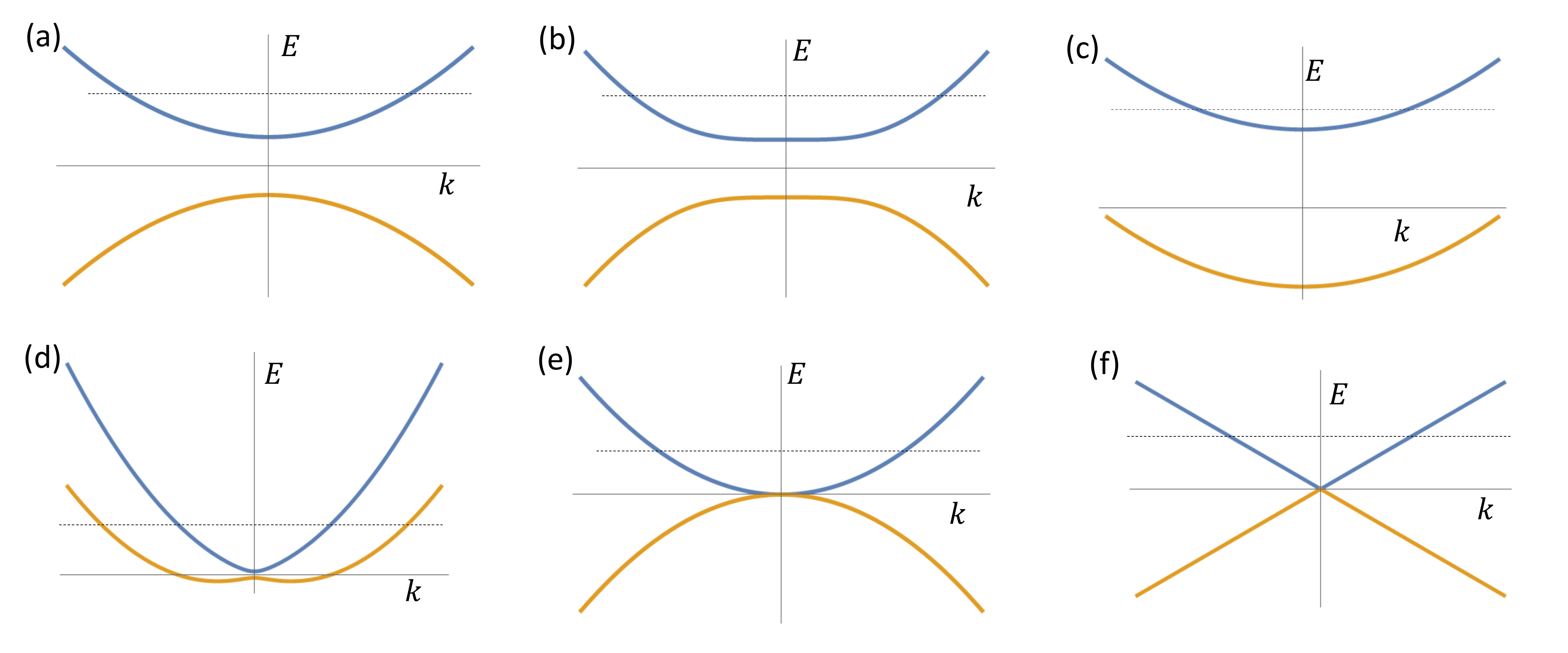}
  \caption{\label{Fig:Band_schematic}
  Schematic band dispersions for various cases from the $\vec{k}\cdot \vec{p}$ Hamiltonian in Eq.~\ref{Eq:k-pHam}:
  (a) $v=0$, $\lambda, 1/m\neq 0 $, and $\tilde{s} = -$, (b) $1/m = 0$, $\lambda, v\neq 0$, and $\gamma = 2$, (c) $v=0$, $\lambda, 1/m\neq 0 $,  and $\tilde{s} = +$, (d) $\lambda,\, v, \, 1/m \neq 0$, and $\tilde{s} = +$, (e) $1/m, \lambda = 0$, $ v \neq 0 $, and $\gamma = 2$,  and (f) $1/m, \lambda = 0$, $ v \neq 0 $, and $\gamma = 1$.
  The black dashed line represents a given position of the chemical potential.
  The cases discussed explicitly in the main text correspond to (a) and (b), while (c) resembles the Zeeman-split case discussed below corresponding to valley-triplet, spin-singlet pairs, and (d) represents a more general case where both bands are present at the Fermi level. 
}
\end{figure}

We are interested in evaluating the ladder diagrams shown in Fig.~\ref{Fig:Ladder_diagram} for a $\delta(\vec{r})$ function interaction.
Since the C.O.M. guiding center $Y_c$ of the pair is conserved during the scattering event, the ladder sum in the Landau gauge basis takes the form
\begin{align}\label{Eq:Supp_scattering}
     & \Gamma(\vec{r}_1,\vec{r}_2,\vec{r}'_1,\vec{r}'_2:i\omega) = \sum_{Y_c} \sum_{\tau}\sum_{\vec{N},\vec{M},Y_r}\sum_{\vec{N}',\vec{M}',Y_{r'}} \Gamma(\vec{N},\vec{M},Y_r,\tau;\vec{N}',\vec{M}',Y'_r,\tau:i\omega) \notag\\
     &\hspace{4cm}\times
     \psi^{\tau}_{\vec{N},Y_c+\frac{Y_r}{2}}(\vec{r}_1)\psi^{-\tau}_{\vec{M},Y_c-\frac{Y_r}{2}}(\vec{r}_2)\psi^{\dagger,\tau}_{\vec{N}',Y_c+\frac{Y_{r'}}{2}}(\vec{r}'_1)\psi^{\dagger,-\tau}_{\vec{M}',Y_c-\frac{Y_{r'}}{2}}(\vec{r}'_2)\, .
\end{align}

We make the assumption that the effective interaction is independent of frequency and the ladder sum reduces to
\begin{align}\label{Eq:Supp_ladderSum}
    &\Gamma(\vec{N},\vec{M},Y_r,\tau;\vec{N}',\vec{M}',Y'_r,\tau:i\omega) =   \biggl \langle \vec{N},\frac{Y_r}{2},\tau;\vec{M},-\frac{Y_r}{2},-\tau\biggl |\, \hat{V}(\vec{r})\, \biggr |\vec{N}',\frac{Y'_r}{2},\tau;\vec{M}',-\frac{Y'_r}{2},-\tau \biggr \rangle\notag\\
     &\hspace{4cm}+ \sum_{\substack{\vec{N}'',\vec{M}'',\\Y''_r}} \biggl [ \mathcal{K}^{\tau}_{\vec{N}'',\vec{M}''}(i\omega) \biggl \langle \vec{N},\frac{Y_r}{2},\tau;\vec{M},-\frac{Y_r}{2},-\tau\biggl |\, \hat{V}(\vec{r})\, \biggr |\vec{N}'',\frac{Y''_r}{2},\tau;\vec{M}'',-\frac{Y''_r}{2},-\tau\biggr \rangle\notag\\
     &\hspace{10cm}\times\Gamma (\vec{N}'',\vec{M}'',Y''_r,\tau;\vec{N}',\vec{M}',Y'_r,\tau:i\omega)\,\biggr ] .
\end{align}

Physically, Eq.~\ref{Eq:Supp_ladderSum} represents two incoming electrons with LL indices $\vec{N}', \vec{M}'$ in opposite valleys scattering via  interaction $\hat{V}(\vec{r})$ to LL indices $\vec{N}, \vec{M}$.
Since the interaction only depends on the relative coordinate, the interaction vertex leaves the C.O.M. guiding center ($Y_c$) invariant.
The first term on the right-hand side of Eq.~\ref{Eq:Supp_ladderSum} is the matrix element of the effective electron-electron interaction, and
\begin{align}\label{Eq:Supp_Kern}
     & \mathcal{K}^{\tau}_{\vec{N},\vec{M}}(i\omega) = \frac{\tanh(\beta (E^{\tau}_{\vec{N}}-\mu)/2)+\tanh(\beta (E^{-\tau}_{\vec{M}}-\mu)/2)}{2(i\omega-E^{\tau}_{\vec{N}}-E^{-\tau}_\vec{M}+2\mu)}\, 
\end{align}
is the two-particle Green's function of free electrons with $\mu$ the chemical potential.

Next, we perform the series sum to rewrite the equation in a more convenient form
\begin{align}\label{Eq:Supp_ladderSeries}
    &\Gamma(\vec{N},\vec{M},Y_r,\tau;\vec{N}',\vec{M}',Y'_r,\tau;i\omega) =   \biggl \langle \vec{N},\frac{Y_r}{2},\tau;\vec{M}-\frac{Y_r}{2},-\tau\biggl |\, \hat{V}(\vec{r}) (\mathbb{1}-\hat{A})^{-1}\, \biggr |\vec{N}',\frac{Y'_r}{2},\tau;\vec{M}',-\frac{Y'_r}{2},-\tau \biggr \rangle \, ,
\end{align}
where
\begin{align}\label{Eq:Supp_scatterng_matrix}
    \hat{A} = \sum_{\vec{N},\vec{M},Y_r,\tau} \mathcal{K}^{\tau}_{\vec{N},\vec{M}}(i\omega)\biggl |\vec{N},\frac{Y_r}{2},\tau;\vec{M},-\frac{Y_r}{2},-\tau\biggl \rangle \biggr \langle \vec{N},\frac{Y_r}{2},\tau;\vec{M},-\frac{Y_r}{2},-\tau\biggl | \hat{V}(\vec{r})\, .
\end{align}

\begin{figure}[!htb]
  \includegraphics[width=0.8\textwidth]{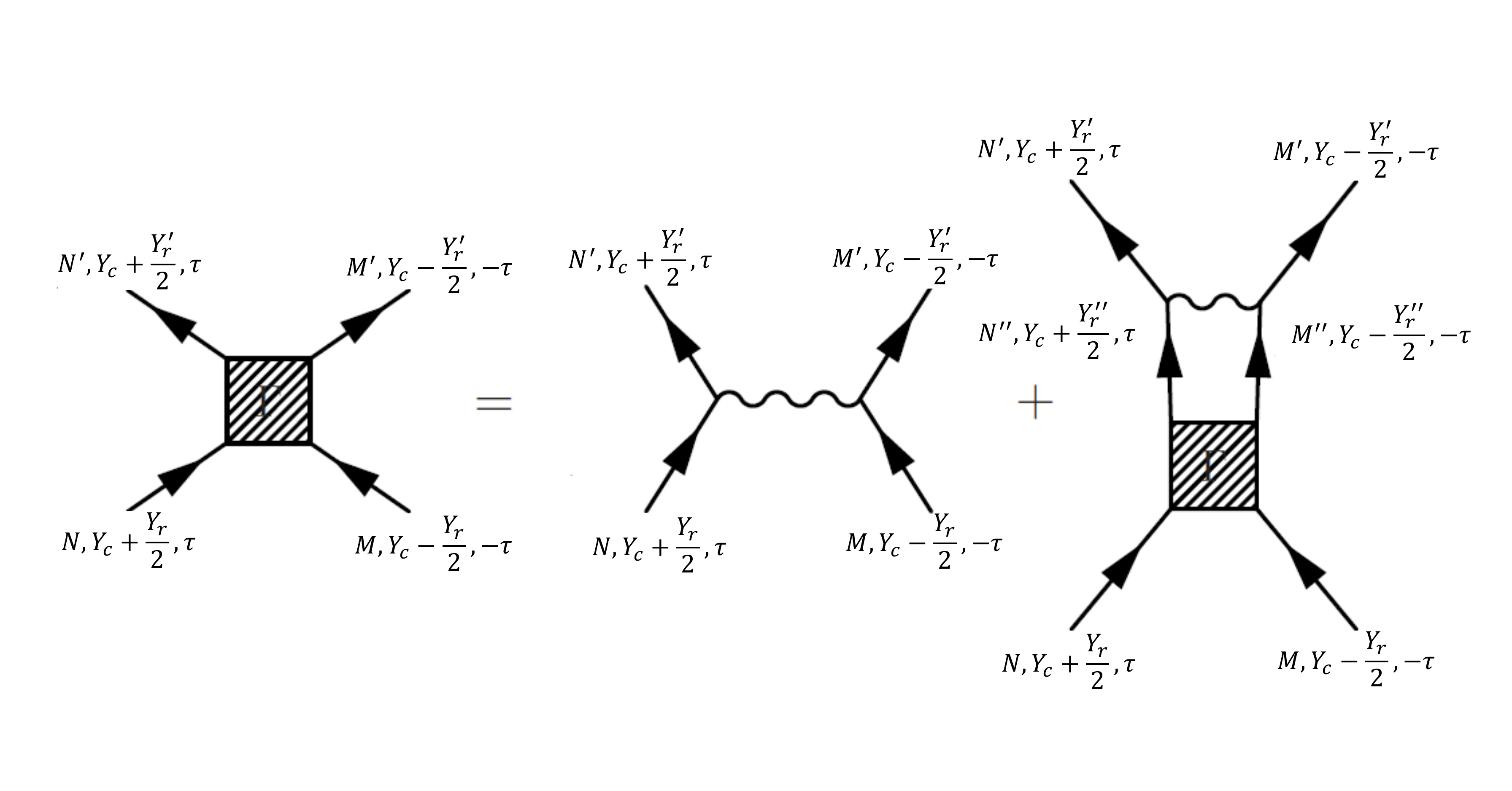}
  \caption{\label{Fig:Ladder_diagram}
  The ladder diagram for the particle-particle scattering function in the LL  representation, with the wavy line representing the pairing interaction.}
\end{figure}

Now, we evaluate the matrix elements in Eq.~\ref{Eq:Supp_ladderSeries}. To do so, we first write down an important integral identity which will be used in the rest of the analysis
\begin{align}\label{EqLSupp_integral}
    & \int d\vec{r}_1 d\vec{r}_2 \,   \phi^{\ast}_{N',Y_c+\frac{Y'_r}{2}}(\vec{r}_1)\phi^{\ast}_{M',Y_c-\frac{Y'_r}{2}}(\vec{r}_2)   \delta(\vec{r}_1-\vec{r}_2)\phi_{N,Y_c+\frac{Y_r}{2}}(\vec{r}_1)\phi_{M,Y_c-\frac{Y_r}{2}}(\vec{r}_2)\notag\\
    &\hspace{8cm} = \frac{1}{L_x}\sum_{j} \mathcal{B}^{N',M'}_j\mathcal{B}^{N,M}_j \varphi^{r}_{N'+M'-j}(-Y'_r)\varphi^r_{N+M-j}(-Y_r)\, ,
\end{align}
which can be obtained by using C.O.M. and relative transformation identities in Eq.~\ref{Eq:Supp_Transformation}.

For simplicity, we consider the two separate cases of purely intra-orbital or purely inter-orbital interactions. 
Under this simplification, the pair wavefunction in the LL basis is effectively a two component wavefunction. 
In what follows we take an effective short range interaction of the form
\begin{align}\label{Eq:Supp_interaction}
    \hat{V}(\vec{r}) \equiv \hat{V}\delta(\vec{r}) = \begin{pmatrix}
    V_0 & V_1\\
    V_1 & V_0
    \end{pmatrix}\delta(\vec{r})\, ,
\end{align}
where the matrix refers to the orbital space.

i) Intra-orbital interaction:
When the short-range interaction is purely intra-orbital, the interaction matrix elements after transforming to C.O.M. and relative coordinates lead to
\begin{align}\label{Eq:Supp_interaction_AA}
    & \biggl \langle \vec{N},\frac{Y_r}{2},\tau;\vec{M},-\frac{Y_r}{2},-\tau \biggl |\, \hat{V}(\vec{r})\, \biggr |\vec{N}',\frac{Y'_r}{2},\tau;\vec{M}',-\frac{Y'_r}{2},-\tau \biggr \rangle = -\frac{1}{L_x}\sum_j (V_0\mathcal{P}^{\vec{N},\vec{M}}_{j,\tau}\mathcal{P}^{\vec{N}',\vec{M}'}_{j,\tau}+V_0\mathcal{Q}^{\vec{N},\vec{M}}_{j,\tau}\mathcal{Q}^{\vec{N}',\vec{M}'}_{j,\tau} \notag\\
    &\hspace{5cm} + V_1 \mathcal{P}^{\vec{N},\vec{M}}_{j,\tau}\mathcal{Q}^{\vec{N}',\vec{M}'}_{j,\tau} + V_1 \mathcal{Q}^{\vec{N},\vec{M}}_{j,\tau}\mathcal{P}^{\vec{N}',\vec{M}'}_{j,\tau})\varphi^{r}_{N+M-\gamma-j}(-Y_{r})\varphi^{r}_{N'+M'-\gamma-j}(-Y'_{r})\, .
\end{align}
Above, the coefficients $\mathcal{P}^{\vec{N},\vec{M}}_{j,\tau}$ and $\mathcal{Q}^{\vec{N},\vec{M}}_{j,\tau}$ are particular combinations of the LL diagonalization coefficients and the C.O.M.-relative transformation coefficients, explicitly given by
\begin{align}\label{Eq:Supp_coeff1}
    & \mathcal{P}^{\vec{N},\vec{M}}_{j,\tau} = \alpha^\tau_{\vec{N}}\beta^{-\tau}_{\vec{M}}\mathcal{B}^{N-\gamma,M}_{j}\, ,\quad \mathcal{Q}^{\vec{N},\vec{M}}_{j,\tau} = \beta^{\tau}_{\vec{N}}\alpha^{-\tau}_{\vec{M}}\mathcal{B}^{N,M-\gamma}_{j}\ ,
\end{align}
and $V_0$ and $V_1$ are respectively the amplitudes of an intra-orbital pair to scatter to the same or to the opposite orbital after an interaction event.

The separation of the interaction matrix elements in different C.O.M. channels $j$ after some algebra leads us to the final evaluation of the ladder sum
\begin{align}\label{Eq:Supp_final_sum}
    & \Gamma(\vec{N},\vec{M},Y_r,\tau;\vec{N}',\vec{M}',Y'_r,\tau:i\omega) = -\frac{1}{L_x}\sum_j \mathcal{A}^{-1}_{\tau,j}  \varphi^{r}_{N+M-\gamma-j}(-Y_{r})\varphi^{r}_{N'+M'-\gamma-j}(-Y'_{r})\, ,
\end{align}
where
\begin{align}\label{Eq:Supp_final_sum2}
    & \mathcal{A}^{-1}_{\tau, j} = \frac{1}{(1-X^{\tau}_j)(1-X^{-\tau}_j)-W^{\tau}_j W^{-\tau}_j}\biggl (  [V_0 (1-X^{-\tau}_j)+V_1 W^{\tau}_j ]\mathcal{P}^{\vec{N},\vec{M}}_{j,\tau}\mathcal{P}^{\vec{N}',\vec{M}'}_{j,\tau} + [V_1 (1-X^{\tau}_j)+ V_0 W^{-\tau}_j ] \mathcal{Q}^{\vec{N},\vec{M}}_{j,\tau}\mathcal{P}^{\vec{N}',\vec{M}'}_{j,\tau} \notag\\
    &\hspace{4cm}+  [ V_0(1-X^{\tau}_j) + V_1W^{-\tau}_j ] \mathcal{Q}^{\vec{N},\vec{M}}_{j,\tau}\mathcal{Q}^{\vec{N}',\vec{M}'}_{j,\tau} + [ V_1(1-X^{-\tau}_j)+V_0W^{\tau}_j ] \mathcal{P}^{\vec{N},\vec{M}}_{j,\tau}\mathcal{Q}^{\vec{N}',\vec{M}'}_{j,\tau} \biggr )\, 
\end{align}
and 
\begin{subequations}
\begin{align}\label{Eq:Supp_final_sum3}
    & X^{\tau}_j = -\frac{1}{4\pi\ell^2 }\sum_{\vec{N},\vec{M}} \mathcal{K}^{\tau}_{\vec{N},\vec{M}}(i\omega)\mathcal{P}^{\vec{N},\vec{M}}_{j,\tau}(V_0 \mathcal{P}^{\vec{N},\vec{M}}_{j,\tau} + V_1 \mathcal{Q}^{\vec{N},\vec{M}}_{j,\tau})\, , \\
    & W^{\tau}_j = -\frac{1}{4\pi\ell^2 }\sum_{\vec{N},\vec{M}} \mathcal{K}^{\tau}_{\vec{N},\vec{M}}(i\omega)\mathcal{P}^{\vec{N},\vec{M}}_{j,\tau}(V_1 \mathcal{P}^{\vec{N},\vec{M}}_{j,\tau} + V_0 \mathcal{Q}^{\vec{N},\vec{M}}_{j,\tau})\, .
\end{align}
\end{subequations}
Here, we have suppressed the LL indices in $\mathcal{A}_{\tau,j}$ and used the relations $\mathcal{K}^{\tau}_{\vec{N},\vec{M}} = \mathcal{K}^{-\tau}_{\vec{M},\vec{N}} $ and $|\mathcal{P}^{\vec{N},\vec{M}}_{j,\tau}| = |\mathcal{Q}^{\vec{M},\vec{N}}_{j,-\tau}|$ in obtaining the final expression in Eq.~\ref{Eq:Supp_final_sum2}.

From Eq.~\ref{Eq:Supp_final_sum2}, the ladder sum diverges when
\begin{align}\label{Eq:Supp_instability}
   & (1-X^{\tau}_j)(1-X^{-\tau}_j) - W^{\tau}_j W^{-\tau}_j = 1\, .
\end{align}
In the limit $V_0 = V_1$, using Eq.~\ref{Eq:Supp_final_sum3}, we also get $X^{\tau}_j = W^{\tau}_j$.  
Finally, taking $\gamma=2$ in the $\omega\rightarrow 0$ limit, we obtain  Eq.~\ref{Eq:linearizedTc1} of the main text from Eq.~\ref{Eq:Supp_instability}.

ii) Inter-orbital interaction:
When the short-range interaction is purely inter-orbital, the interaction matrix elements after transforming to C.O.M. and relative coordinates lead to a sum over C.O.M. channels
\begin{align}\label{Eq:Supp_interaction_AB}
    & \biggl \langle \vec{N},\frac{Y_r}{2},\tau;\vec{M},-\frac{Y_r}{2},-\tau \biggl |\, \hat{V}(\vec{r})\, \biggr |\vec{N}',\frac{Y'_r}{2},\tau;\vec{M}',-\frac{Y'_r}{2},-\tau \biggr \rangle
    = -\frac{1}{L_x}\sum_j (\Psi^{\vec{N},\vec{M},\tau}_j)^T \hat{V} \Psi^{\vec{N}',\vec{M}',\tau}_j\
\end{align}
where
\begin{align}\label{Eq:Supp_interaction_AB1}
    \Psi^{\vec{N},\vec{M},\tau}_j = \biggl (
    \mathcal{R}^{\vec{N},\vec{M}}_{j,\tau} \varphi^r_{N+M-2\gamma-j}(-Y_r),\, 
    \mathcal{S}^{\vec{N},\vec{M}}_{j,\tau} \varphi^r_{N+M-j}(-Y_r)
    \biggr )^T\,
\end{align}
and
\begin{align}\label{Eq:Supp_coeff2}
    & \mathcal{R}^{\vec{N},\vec{M}}_{j,\tau} = \alpha^\tau_{\vec{N}}\alpha^{-\tau}_{\vec{M}}\mathcal{B}^{N-\gamma,M-\gamma}_{j}\, ,\quad \mathcal{S}^{\vec{N},\vec{M}}_{j,\tau} = \beta^\tau_{\vec{N}}\beta^{-\tau}_{\vec{M}}\mathcal{B}^{N,M}_{j}\
\end{align}
are the transformation coefficients. Similar to the intra-orbital case, $V_0$ and $V_1$ are respectively the amplitudes of an inter-orbital pair to preserve or flip the orbital index of the individual electron during the scattering event.

The ladder sum in this case mixes different C.O.M. channels, hence we are unable to obtain a compact expression for the full ladder sum. 
However, we can identify relevant pairing channels that still diverge for different C.O.M. channels. To see that, we represent the $n^{\text{th}}$ order scattering contribution as
\begin{align}\label{Eq:Supp_nth_interaction}
    & \biggl \langle \vec{N},\frac{Y_r}{2},\tau;\vec{M},-\frac{Y_r}{2},-\tau \biggl |\, \hat{V} (\vec{r})\, \hat{A}^n\biggr |\vec{N}',\frac{Y'_r}{2},\tau;\vec{M}',-\frac{Y'_r}{2},-\tau \biggr \rangle \sim -\frac{1}{L_x} \sum_{j}   (\Psi^{\vec{N},\vec{M},\tau}_j)^T  \hat{V} (\hat{U}^{n-1}_{\tau,j}+ \hat{O}^{n} ) \Psi^{\vec{N}',\vec{M}',\tau}_j\,
\end{align}
where
\begin{align}\label{Eq:Supp_nth_interaction1}
    & \hat{U}_{\tau,j} = -\frac{1}{4\pi\ell^2}\begin{pmatrix}
    \sum_{\vec{N},\vec{M}} \mathcal{K}^{\tau}_{\vec{N},\vec{M}} (\mathcal{R}^{\vec{N},\vec{M}}_{j,\tau})^2 & 0\\
    0 & \sum_{\vec{N},\vec{M}} \mathcal{K}^{\tau}_{\vec{N},\vec{M}} (\mathcal{S}^{\vec{N},\vec{M}}_{j,\tau})^2
    \end{pmatrix}\hat{V}\, ,
\end{align}
and $\hat{O}^n$ is an additional term. Here in Eq.~\ref{Eq:Supp_nth_interaction}, we have ignored extra channel mixing  terms that do not lead to an infinite series.  
The lowest non-zero contribution from the additional term $\hat{O}^n$ considered in Eq.~\ref{Eq:Supp_nth_interaction}  appears at third order in the scattering.
Similarly, after lengthy algebra one can write down higher order terms, however their general form is very complicated to write down here. For our arguments below, their exact expressions are not relevant. 

Next, we can represent the ladder sum as 
\begin{align}\label{Eq:Supp_final_sum_inter}
    & \Gamma(\vec{N},\vec{M},Y_r,\tau;\vec{N}',\vec{M}',Y'_r,\tau:i\omega) \sim -\frac{1}{L_x}\sum_j  (\Psi^{\vec{N},\vec{M},\tau}_j)^T \hat{V} \biggl [ (\mathbb{1}- \hat{U}_{\tau,j})^{-1}  + \sum_n \hat{O}^{n}  \biggr ] \Psi^{\vec{N}',\vec{M}',\tau}_j \, .
\end{align}
Since we are only interested in the pairing instability, we need to look at the divergences in the above equation. One clear source of divergence is when the $2\times 2$ matrix $(\mathbb{1}-\hat{U}_{j,\tau})^{-1}$ becomes singular, which leads to the condition
\begin{align}\label{Eq:Supp_divergence_inter}
    & \frac{V_0}{4\pi\ell^2} \sum_{\vec{N},\vec{M}} \mathcal{K}^{\tau}_{\vec{N},\vec{M}} [(\mathcal{R}^{\vec{N},\vec{M}}_{j,\tau})^2+(\mathcal{S}^{\vec{N},\vec{M}}_{j,\tau})^2]  + \frac{V^2_1-V^2_0}{(4\pi\ell^2)^2}\biggl [ \sum_{\vec{N},\vec{M}}  \mathcal{K}^{\tau}_{\vec{N},\vec{M}} (\mathcal{R}^{\vec{N},\vec{M}}_{j,\tau})^2 \biggr]\biggl [ \sum_{\vec{N},\vec{M}}  \mathcal{K}^{\tau}_{\vec{N},\vec{M}} (\mathcal{S}^{\vec{N},\vec{M}}_{j,\tau})^2 \biggr]   = 1\, .
\end{align}
When the above equation is satisfied for any pairing channel $j$, the only way to cancel the divergence is if the sum $\sum_n \hat{O}^n$ contributes a negative divergence. 
However the contribution of the pair mixing term is non-negative definite, hence it cannot cancel the divergence.
Thus, we conclude that Eq.~\ref{Eq:Supp_divergence_inter} indeed leads to a pairing instability. 
Finally, taking the $V_0 = V_1$ limit, we arrive at Eq.~\ref{Eq:linearizedTc2} of the main text.

\section{Ladder sum for valley-triplet and spin-singlet pairs\label{Sec:spin_singlet}}
So far, we have assumed spin-triplet pairs. However, since there is still debate on the actual nature of the pairs, we also discuss spin-singlet pairs here.
In this context, Ref.~\cite{Das2021} indicates broken LL degeneracy for MATBG.
This is possible if the active bands are Zeeman split.
In this case, the possible pairs might be valley-triplet and spin-singlet.
We can similarly calculate the ladder sum for the spin-singlet pairs. 
Since in this case, the pairs are intravalley, we drop the valley $\tau$ index.
Further the Hamiltonian in Eq.~\ref{Eq:supp_magneticHam} should be considered in the spin basis and then $\lambda$ is a Zeeman energy term.
The ladder sum takes the form
\begin{align}\label{Eq:Supp_ladderSumSpin}
    &\Gamma(\vec{N},\vec{M},Y_r;\vec{N}',\vec{M}',Y'_r:i\omega) =   \biggl \langle \vec{N},\frac{Y_r}{2};\vec{M},-\frac{Y_r}{2}\biggl |\, \hat{V}(\vec{r})\, \biggr |\vec{N}',\frac{Y'_r}{2};\vec{M}',-\frac{Y'_r}{2} \biggr \rangle\notag\\
     &+ \sum_{\substack{\vec{N}'',\vec{M}'',\\Y''_r}} \mathcal{K}_{\vec{N}'',\vec{M}''}(i\omega) \biggl \langle \vec{N},\frac{Y_r}{2};\vec{M},-\frac{Y_r}{2}\biggl |\, \hat{V}(\vec{r})\, \biggr |\vec{N}'',\frac{Y''_r}{2};\vec{M}'',-\frac{Y''_r}{2}\biggr \rangle \, \Gamma (\vec{N}'',\vec{M}'',Y''_r;\vec{N}',\vec{M}',Y'_r:i\omega)\, ,
\end{align}
where $\mathcal{K}_{\vec{N},\vec{M}}$ has the same form as in Eq.~\ref{Eq:Supp_Kern} after dropping the $\tau$ index.
After the series sum, this reduces to
\begin{align}\label{Eq:Supp_ladderSeriesSpin}
    &\Gamma(\vec{N},\vec{M},Y_r;\vec{N}',\vec{M}',Y'_r;i\omega) =   \biggl \langle \vec{N},\frac{Y_r}{2};\vec{M}-\frac{Y_r}{2}\biggl |\, \hat{V}(\vec{r}) (\mathbb{1}-\hat{A})^{-1}\, \biggr |\vec{N}',\frac{Y'_r}{2};\vec{M}',-\frac{Y'_r}{2} \biggr \rangle \, ,
\end{align}
where $\hat{A}$ also has the same form as in Eq.~\ref{Eq:Supp_scatterng_matrix} after dropping the $\tau$ index.
The final evaluation of the ladder sum is
\begin{align}\label{Eq:Supp_final_sumSpin}
    & \Gamma(\vec{N},\vec{M},Y_r;\vec{N}',\vec{M}',Y'_r:i\omega) = -\frac{1}{L_x}\sum_j \mathcal{A}^{-1}_{j}  \varphi^{r}_{N+M-\gamma-j}(-Y_{r})\varphi^{r}_{N'+M'-\gamma-j}(-Y'_{r})\, ,
\end{align}
where $\mathcal{A}_j$ is the same as Eq.~\ref{Eq:Supp_final_sum2}.
So, we again obtain the instability relation Eq.~\ref{Eq:Supp_instability}.

\begin{figure}[!htb]
  \includegraphics[width=0.45\textwidth]{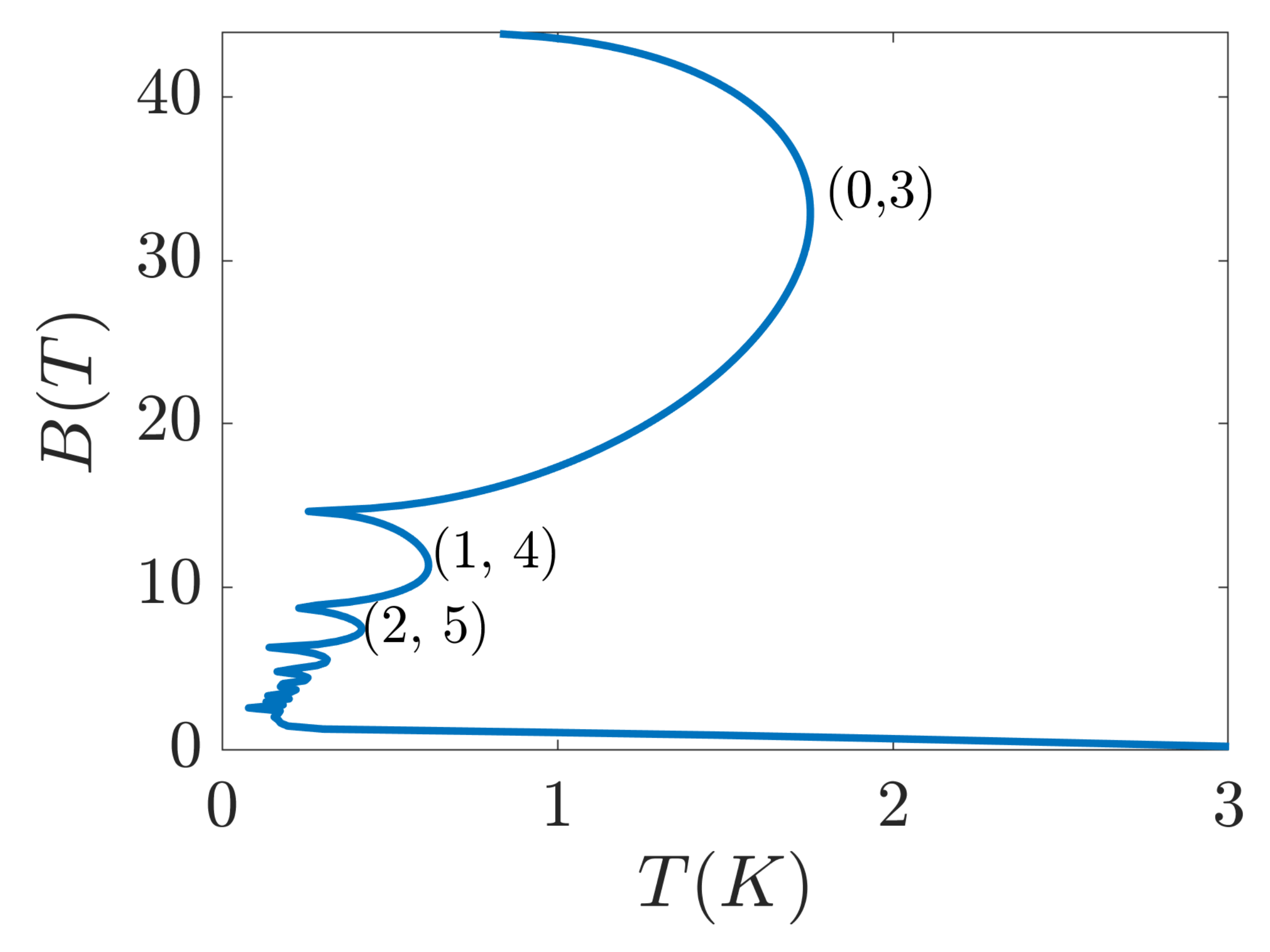}
  \caption{\label{Fig:supp_spinSinglet}
   Magnetic field versus temperature phase diagram assuming spin-singlet pairing and a g-factor of 2. The field is tilted such that the majority spin LL $N$ and the minority spin LL $N+3$ are degenerate, which corresponds to a tilt angle that satisfies $\cos(\theta)=m^\ast/m_e$.
  }
\end{figure}

To demonstrate pairing in the spin-singlet and valley-triplet case, we take $\tilde{s} = +$ and the $v\rightarrow 0$ limit.
The two bands in the single valley then resemble Fig.~\ref{Fig:Band_schematic}c.  
We can associate the upper band of the Hamiltonian in Eq.~\ref{Eq:Supp_kp} with the minority spin and the lower band with the majority spin.
Even though we have taken the $v\rightarrow 0$ limit, to be consistent with the discussion in the main text  we take $\gamma = 2$, which leads to a LL counting similar to the main text.
The band gap parameter $\lambda$ resembles a Zeeman energy, hence we replace $\lambda = g\mu_B B$, where $g$ is an effective $g$-factor. We introduce this Zeeman energy shift in the two energy arguments of $\mathcal{K}_{\vec{N},\vec{M}}$, these being opposite for the spin-up and spin-down levels.  
Any finite g-factor will lead to a rapid suppression of $T_c$ in the quantum regime~\cite{Norman1990}.

This can be ameliorated by tilting the magnetic
field in order to bring a pair of LLs into degeneracy, though $T_c$ will be suppressed because of the difference in LL index between the
up and down spin levels. 
For a parabolic dispersion, this can be achieved if the tilt angle satisfies $\cos \theta = gm^{\ast}/(2m_e)$, in which case the LL indices differ by three (a difference of two because of the LL counting difference of two between the upper and lower bands in our notation, and an additional difference of one introduced by the Zeeman shift).
In Fig.~6, we show that reentrant superconductivity  can still exist for spin-singlet pairs, though with a suppressed $T_c$.
It is likely that the actual pair of active bands are some emergent spin-valley flavor~\cite{Park2021a}.
We expect our qualitative result of robust reentrant superconductivity to still be true in that case.

\end{widetext}


\bibliography{bibliography}

\end{document}